\documentclass[12pt]{article}
\usepackage{graphics,epsfig}
\newcounter{saveeqn}

\begin{document}

\begin{titlepage}\hfill HD-THEP-03-22\\[5 ex]

\begin{center}{\Large\bf On the Glue Content in Heavy Quarkonia} \\[5 ex]

{\large \bf Dieter Gromes}\\[3 ex]Institut f\"ur
Theoretische Physik der Universit\"at Heidelberg\\ Philosophenweg 16,
D-69120 Heidelberg \\ E - mail: d.gromes@thphys.uni-heidelberg.de \\
 \end{center} \vspace{2cm}

{\bf Abstract:} Starting with two coupled Bethe-Salpeter equations for the quark-anti\-quark, and for the quark-glue-antiquark component of the quarkonium, we solve the bound state equations perturbatively. The resulting admixture of glue can be partially understood in a semiclassical way, one has, however, to take  care of the different use of time ordered versus retarded Green functions. Subtle questions concerning the precise definition of the equal time wave function arise, because the wave function for the Coulomb gluon is discontinuous with respect to the relative time of the gluon. A striking feature is that a one loop non abelian graph contributes to the same order as tree graphs, because the couplings of transverse gluons in the tree graphs are suppressed in the non relativistic bound state, while the higher order loop graph can couple to quarks via non suppressed Coulomb gluons. We also calculate the amplitude for quark and antiquark at zero distance in the quark-glue-antiquark component of the $P$-state. This quantity is of importance for annihilation decays of $P$-states. It shows a remarkable compensation between the tree graph and the non abelian loop graph contribution. An extension of our results to include non perturbative effects is possible.

 \vfill \centerline{April  2003}

\end{titlepage}

\section{Introduction}

It is generally agreed that a quarkonium state is not only made up  of a quark and an antiquark ($Q\bar{Q}$), but contains also a component with glue ($QA\bar{Q}$). There are, in addition,  other contributions like admixtures of several gluons or light quark-antiquark pairs which are not considered here. We emphasize that we are not talking about hybrids, where the glue component is an essential part of the wave function, but about the admixture of glue in ordinary quarkonia. We are mainly interested in equal time wave functions here, not, say, in light cone wave functions.  

The presence of a component with glue in the $P$-states has become of great importance due to the work of Bodwin, Braaten, and Lepage \cite{BBL}. Although this component is suppressed, it allows the $Q\bar{Q}$ pair in the $QA\bar{Q}$ component to be in a color octet $S$-state. The $S$-state  annihilation of this pair is of the same order as the $Q\bar{Q}$ annihilation in the leading $Q\bar{Q}$ component of the $P$-state, because the latter is also suppressed by the vanishing of the wave function at the origin. Logarithmic divergences cancel in both contributions. Similar considerations hold for $P$-wave production.

In spite of the principal as well as phenomenological importance of the $QA\bar{Q}$ admixture it is hard to find quantitative statements about this part of the wave function in the literature. We will approach the problem in a systematic way here. Even if one is interested in the equal time wave functions only, it is necessary to start with the general wave functions involving different times. The reason is that quark and gluon propagators also depend on time differences, therefore a treatment involving only equal time wave functions appears not possible within our approach. We leave open the question whether our results could also be obtained within an equal time Hamilton approach. In any case some of the subtleties mentioned below can only be understood if one knows the general wave function with different time arguments.

In sect. 2 we use as starting point two Green functions, one involving  $Q\bar{Q}$ and the other one  $QA\bar{Q}$. In the spirit of the usual derivation of the Bethe-Salpeter equation, but with more necessary technicalities, we define several irreducible kernels. These allow a decomposition of the Green functions and a derivation of two coupled Bethe-Salpeter equations for the $Q\bar{Q}$ wave function and the $QA\bar{Q}$ wave function.

In sect. 3 we proceed by using the framework of non relativistic QCD (NRQCD) \cite{NRQCD}, which is appropriate for the case that the velocities of the quarks  are small (order $\alpha _s \ll 1$). We furthermore choose the radiation gauge $\nabla {\bf A}=0$ which is known to be the gauge best suited for treating bound state systems. In leading order of perturbation theory the $Q\bar{Q}$ kernel becomes static and the $QA\bar{Q}$ wave function does not enter into the equation for the $Q\bar{Q}$ wave function. The latter can thus be solved in the usual way. After this has been done one can insert the solution for the $Q\bar{Q}$ wave function into the equation for the $QA\bar{Q}$ wave function and calculate the latter in lowest order. 

The result shows some delicate features. The wave function with the Coulomb gluon is discontinuous at $t=0$, where $t$ is the relative time of the gluon with respect to the time of the $Q\bar{Q}$ system. It vanishes for positive $t$, while for negative $t$ it can be simply understood as the Coulomb field moving along with the quarks. The tree graph contribution to the transverse gluon differs by a factor 1/2 from what one would expect from classical electrodynamics. The origin of this discrepancy can be understood, it is due to the use of time ordered versus retarded propagators in the different approaches. For the transverse gluon there is a non abelian loop contribution, in which the transverse gluon couples to two Coulomb gluons while the latter couple to the quark and antiquark. This loop graph contributes to the same order as the direct coupling of the transverse gluon to quark or antiquark, because the direct coupling is suppressed by a factor of the order of $|{\bf p}|/m \approx \alpha _s$.

In sect. 4 we finally calculate the amplitude for $Q\bar{Q}$ at zero distance in the $QA\bar{Q}$ component of the $P$-state. As stated above, this quantity is of importance for annihilation decays and production of $P$-states. The result cannot simply be written in terms of, say, the derivative of the $Q\bar{Q}$ radial wave function at the origin, it depends on the gluon momentum as well as on the vector index of the gluon and the magnetic quantum number of the $P$-state. 

In the conclusions of sect. 5 we briefly discuss how an extension of our perturbative results to a wider range of applicability can be carried out. Applications to $P$-state annihilation decays need some further effort and will be presented elsewhere.

\setcounter{equation}{0}\addtocounter{saveeqn}{1}%

\section{Coupled Bethe-Salpeter equations}

We consider a mesonic bound state $|K>$ with four momentum $K^\mu $. We are interested both in the quark-antiquark amplitude

\begin{equation} \psi _{Q\bar{Q}}(x_1,x_2;K) = <0|T(\psi (x_1)\bar{\psi }(x_2))|K> =  \frac{e^{-iKX}}{(2\pi)^4}\int e^{-ipr}\tilde{\psi }_{Q\bar{Q}}(p;K)d^4p, \end{equation}
as well as in the quark-glue-antiquark component

\begin{eqnarray} \psi _{QA^\mu\bar{Q}}(x_1,x_3,x_2;K) & = & <0|T(\psi (x_1) A^\mu (x_3)\bar{\psi }(x_2))|K> \nonumber\\ 
 & = & \frac{e^{-iKX}}{(2\pi)^8}\int e^{-i(pr+qx)}\tilde{\psi }_{QA^\mu \bar{Q}}(p,q;K)d^4pd^4q. \end{eqnarray}
Spin- and color indices are suppressed, explicitly the color singlet states are understood as $\psi \bar{\psi }\equiv (1/\sqrt{N}) \bar{\psi }_\alpha \psi _\alpha $, and $\psi  A^\mu \bar{\psi }\equiv\sqrt{2/(N^2-1)}\bar{\psi }_{\alpha _2}  A^{\mu a} (\lambda ^a_{\alpha _2 \alpha _1}/2)\psi _{\alpha _1}$, with $N=3$ the number of colors. There is no need of specifying the gauge at this stage. 

In the two particle $Q\bar{Q}$ system, with masses $m_1,m_2$, we used the usual CM and relative variables in position and momentum space:

\begin{eqnarray} X = \eta _1 x_1 + \eta _2 x_2, \quad r=x_1-x_2 & ; & \; x_1=X+\eta _2 r, \quad x_2=X-\eta _1 r,\nonumber\\
K = p_1 + p_2, \quad p=\eta_2 p_1 - \eta _1 p_2 & ; & \;p_1 = \eta _1 K + p, \quad p_2 = \eta _2 K - p. \end{eqnarray}
Here $\eta _j = m_j/(m_1 + m_2)$, the total and reduced mass are denoted by $M=m_1 + m_2$ and $\mu = m_1  m_2/(m_1 + m_2)$. For the three particle system $QA\bar{Q}$ it is convenient to use the same CM and relative coordinates $X,r$ for $Q$ and $\bar{Q}$ as above. We denote by $x$ the distance of the gluon to the CM of the $Q\bar{Q}$, the momentum of the gluon is denoted by $q$. The coordinates then read

\begin{eqnarray} x = x_3 -\eta _1 x_1 - \eta _2 x_2 & ; & \; x_3 = X+x,\nonumber\\
K = p_1 + p_2+q & ; & \; p_1 = \eta _1 (K-q) + p, \; p_2 = \eta _2 (K-q) - p. \end{eqnarray}
The derivation of the bound state equations for $\tilde{\psi }_{Q\bar{Q}}(p;K)$ and $\tilde{\psi }_{QA^\mu \bar{Q}}(p,q;K)$ is a bit technical. One may skip the rest of this section, the resulting lowest order graphs in Fig. 6 and Fig. 7 of the next section are easily understood anyhow. For a systematic and general approach we consider the Green functions 

\begin{eqnarray} \lefteqn{G_{Q\bar{Q},Q\bar{Q}}(x_1,x_2,x'_1,x'_2)
  =  <0|T([\psi (x_1)\bar{\psi }(x_2)][\psi (x'_1) \bar{\psi }(x'_2)])|0>}\\
& = & \frac{1}{(2\pi)^{12}}\int e^{-i(pr+p'r'+K(X+X'))} \tilde{G}_{Q\bar{Q},Q\bar{Q}}(p,p';K)d^4pd^4p'd^4K, \nonumber\end{eqnarray}
and 

\begin{eqnarray} \lefteqn{ G _{QA^\mu \bar{Q},Q\bar{Q}}(x_1,x_3,x_2,x'_1,x'_2)
  =  <0|T([\psi (x_1)A^\mu(x_3)\bar{\psi }(x_2)][\psi (x'_1)\bar{\psi }(x'_2)])|0>}\\
& = & \frac{1}{(2\pi)^{16}}\int e^{-i(pr+p'r'+qx+K(X+X'))} \tilde{G}_{QA^\mu \bar{Q},Q\bar{Q}}(p,p',q;K)d^4pd^4p'd^4qd^4K. \nonumber \end{eqnarray}
Further Green functions, which will be needed below, are defined in an analogous way and immediately recognized from their symbols. Wave functions and Green functions in momentum space are denoted by open circles with the appropriate legs attached to them. The quark gluon vertex function is defined in the usual way as the sum of all one-particle irreducible graphs contributing to the $QAQ$-vertex, i.e. the graphs which cannot be separated into two different pieces by cutting only one internal line. The external legs are not included. 

With the help of the vertex function the Green function $\tilde{G} _{QA^\mu \bar{Q},Q\bar{Q}}$ can then be decomposed as in Fig. 1, where internal lines, here as well as in the other graphs of this section, denote full propagators.

\begin{figure}[htb]
\begin{center}
\epsfysize 1.6cm
\epsfbox{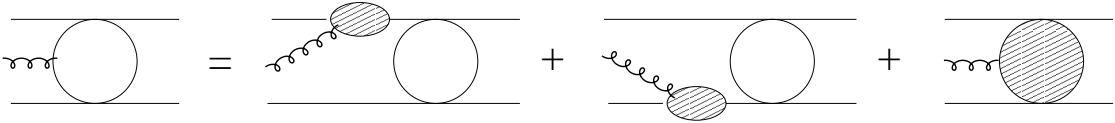}
\end{center}
\caption{ {\it The decomposition of $\tilde{G} _{QA^\mu \bar{Q},Q\bar{Q}}$.}}
\end{figure}    

We define the totally connected Green function $\tilde{G}^{conn}_{QA\bar{Q},Q\bar{Q}}$, denoted with the filled circle in Fig. 1, as the sum of the  graphs which  cannot be divided into two separate pieces by cutting one internal $Q$ (or $\bar{Q}$) line plus the left external $\bar{Q}$ (or $Q$) line.

In order to separate the different components in the bound state we have to introduce  four irreducible kernels $\tilde{I}_{Q\bar{Q},Q\bar{Q}}, \tilde{I}_{Q\bar{Q},QA^\mu \bar{Q}}, \tilde{I}_{QA^\mu \bar{Q},Q\bar{Q}},\tilde{I}_{QA^\mu \bar{Q},QA^\nu \bar{Q}}$. They are defined as the sums of all connected graphs contributing to the corresponding Green functions $\tilde{G}_{Q\bar{Q},Q\bar{Q}}, \cdots ,$ respectively, without external legs, and will be denoted by rectangles with the appropriate lines attached. The irreducible kernels have the following properties: 

\begin{description}
\item[a)] The graphs  cannot be separated into two pieces by cutting two internal lines of the type $Q\bar{Q}$, or by cutting  three internal lines of the type $QA\bar{Q}$. 
\end{description}

The kernels $\tilde{I}_{Q\bar{Q},QA^\mu \bar{Q}}$ and $\tilde{I}_{QA^\mu \bar{Q},QA^\nu \bar{Q}}$  have a $QA\bar{Q}$ state on the right hand side, and  there are three further rules for them: 

\begin{description}
\item[b)] The graph cannot be separated into two pieces by cutting an internal $Q$ (or $\bar{Q}$) plus the right external $\bar{Q}$ (or $Q$) line. 

\item[c)] The graph cannot be separated into two pieces by cutting two internal  $QA$ (or $A\bar{Q}$) lines plus the right external $\bar{Q}$ (or $Q$) line. 

\item[d)] The graph cannot be separated into two pieces by cutting two internal  $Q\bar{Q}$  lines plus the right external $A$ line.
 
The prescriptions b), c), d) only refer to lines on the right hand side. If the corresponding situation appears on the left hand side, the graph is included in the definition of $\tilde{I}_{QA^\mu \bar{Q},Q\bar{Q}}$ or $\tilde{I}_{QA^\mu \bar{Q},QA^\nu\bar{Q} }$. 

\end{description}

The above definitions have been chosen in such a way that one is able to extract the desired wave functions while avoiding double counting in the decompositions below. Note that we do not make use of the usual two quark irreducible kernel $\tilde{I}$ consisting of all connected graphs for $\tilde{G}_{Q\bar{Q},Q\bar{Q}}$ (without external legs) which cannot be separated into two pieces by a $Q\bar{Q}$ cut. We need a more detailed classification of kernels for our purpose.

Consider now the sum of all graphs which contribute to  $\tilde{G}_{Q\bar{Q},Q\bar{Q}}$.  They may be collected in different classes using the above definitions. This is shown in Fig. 2.

\begin{figure}[htb]
\begin{center}
\epsfysize 5.5cm
\epsfbox{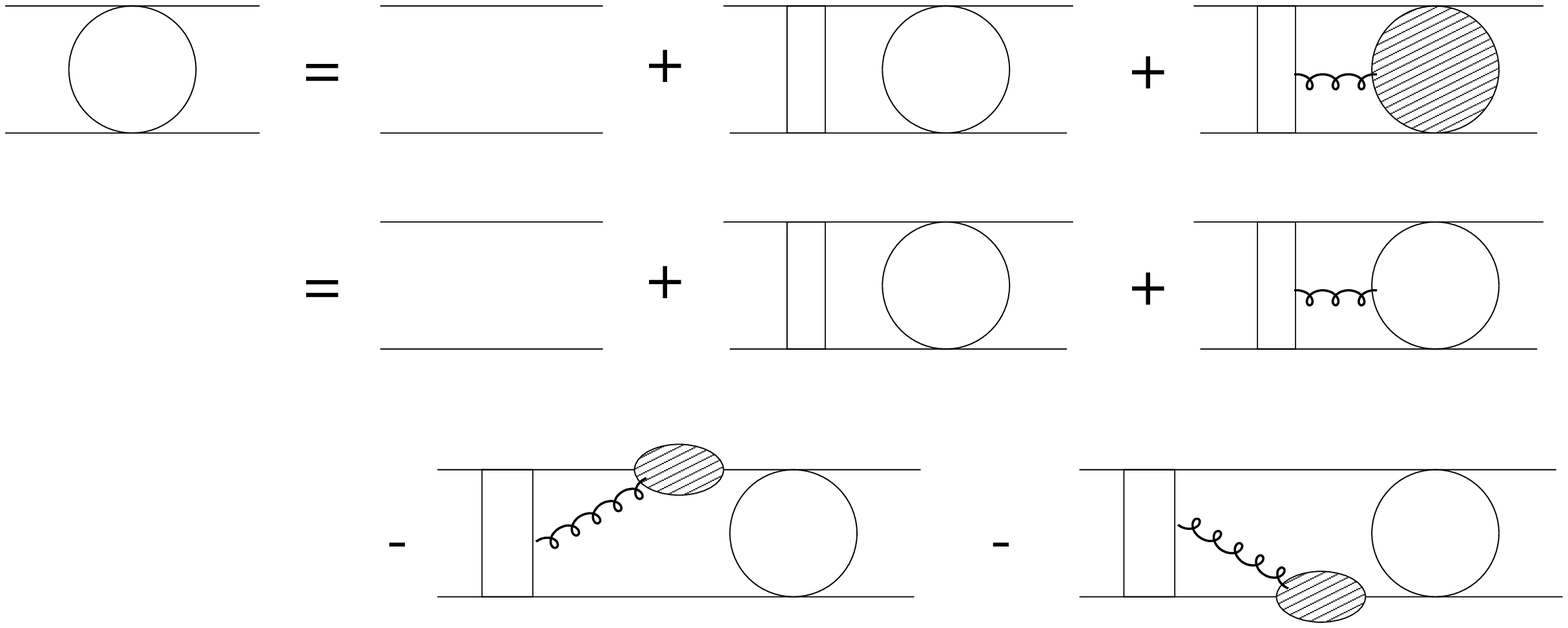}
\end{center}
\caption{ {\it The equation for $\tilde{G}_{Q\bar{Q},Q\bar{Q}}$.}}
\end{figure}    

The corresponding decomposition of $\tilde{G}_{QA^\mu \bar{Q},Q\bar{Q}}$ is shown in Fig. 3.

\begin{figure}[htb]
\begin{center}
\epsfysize 9cm
\epsfbox{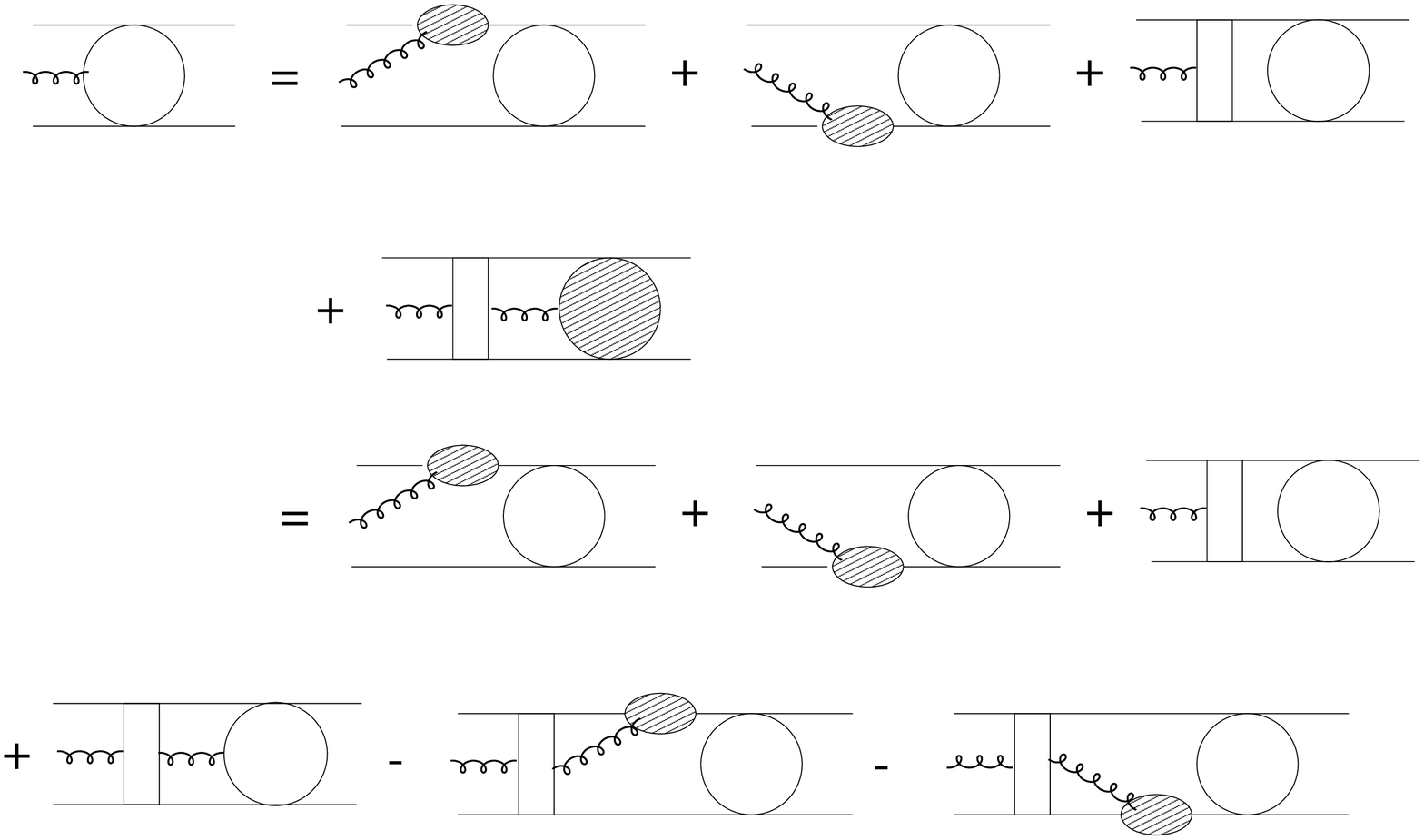}
\end{center}
\caption{ {\it The equation for $\tilde{G}_{QA^\mu \bar{Q},Q\bar{Q}}$.}}
\end{figure}    

The correctness of the decompositions in Fig. 2 and Fig. 3 can be demonstrated in any order of perturbation theory by showing that there is a one to one correspondence between graphs. Clearly each graph appearing on the rhs also belongs to the lhs. The proof that each graph on the lhs appears exactly once on the rhs is given in the appendix. There we also show some examples for graphs which are forbidden by our above definitions.

The intermediate state $|K><K|$ leads to a pole in the Green functions at $K^0 = \pm (\sqrt{M_K^2 + {\bf K}^2} -i\epsilon)$ on both sides of Fig. 2 and Fig. 3. Extracting the residue and dropping the common factor $\sim \tilde{\psi }^\dagger _{Q\bar{Q}}$ we arrive at two coupled Bethe-Salpeter equations for the wave functions $\tilde{\psi }_{Q\bar{Q}}$ and $\tilde{\psi }_{QA^\mu\bar{Q}}$ which are displayed in Fig. 4 and Fig. 5.

\begin{figure}[htb]
\begin{center}
\epsfysize 4cm
\epsfbox{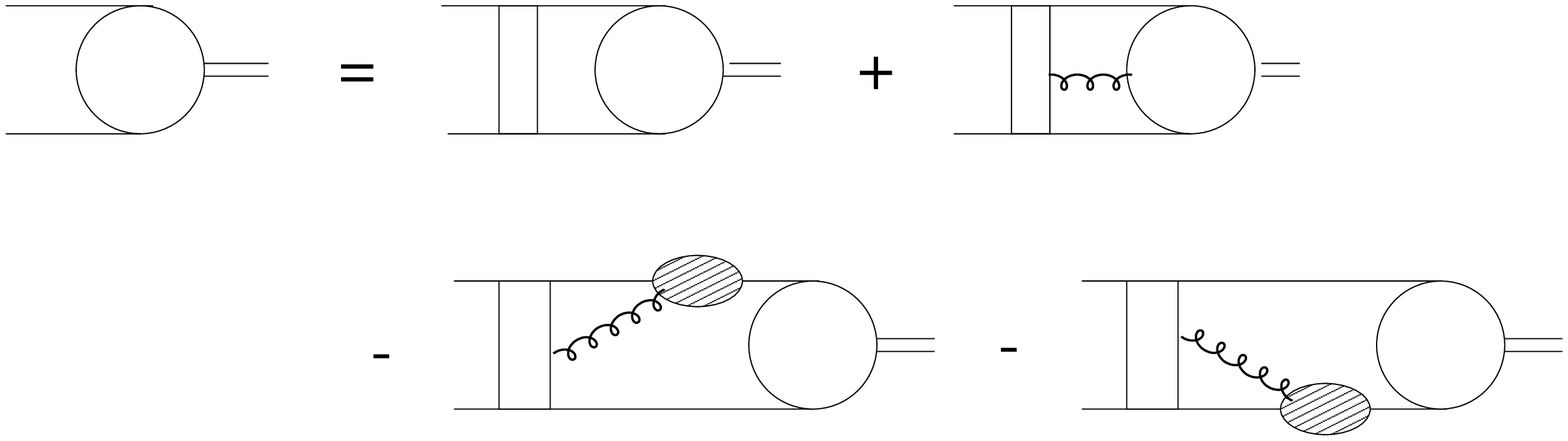}
\end{center}
\caption{ {\it The equation for $\tilde{\psi }_{Q\bar{Q}}$.}}
\end{figure}    

\begin{figure}[htb]
\begin{center}
\epsfysize 3.5cm
\epsfbox{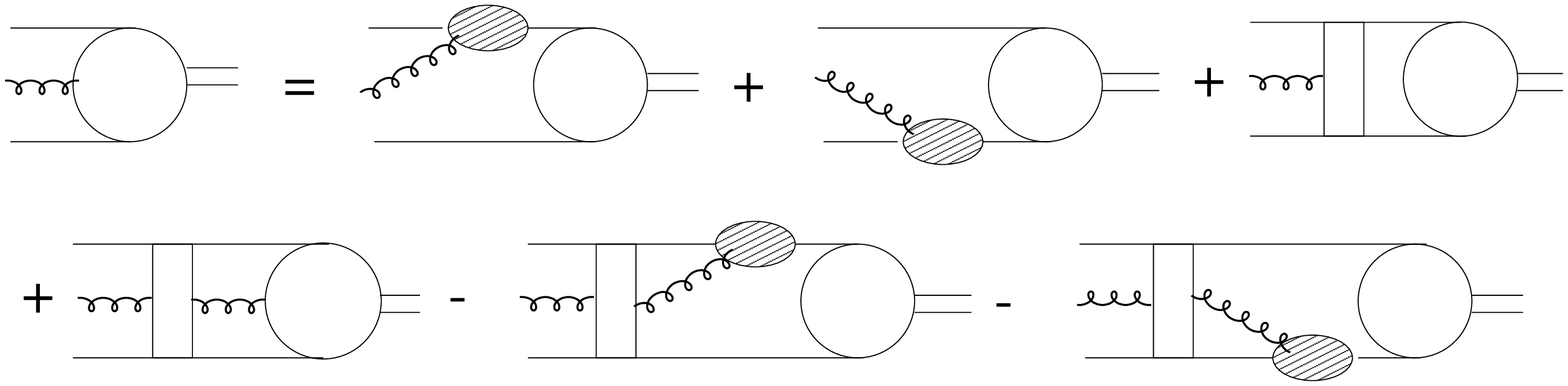}
\end{center}
\caption{ {\it The equation for $\tilde{\psi }_{QA^\mu \bar{Q}}$.}}
\end{figure}    

\setcounter{equation}{0}\addtocounter{saveeqn}{1}%

\section{Perturbative solution}

Up to now the treatment was general and there was no need to specify the Lagrangian or the gauge in detail. In this section we will specialize to non relativistic QCD (NRQCD up to order $1/m$) with the Lagrangian \cite{NRQCD}

\begin{equation} {\cal L}_Q = \psi^+ (iD^0 + \frac{\vec{D}^2}{2m_1} + \frac{g{\bf s}_1{\bf B}}{m_1})\psi + 
\chi^+ (iD^0 - \frac{\vec{D}^2}{2m_2} - \frac{g{\bf s}_2{\bf B}}{m_2})\chi - \frac{1}{2}Tr F^{\mu \nu }F_{\mu \nu }. \end{equation}
Here $\psi $ is the Pauli spinor for the quark field (annihilation operator), and $\chi $ for the antiquark field (creation operator). Our conventions are $D^\mu = \partial ^\mu +igA^\mu,\; F^{\mu \nu} = \partial ^\mu A^\nu - \partial ^\nu A^\mu +ig[A^\mu ,A^\nu ],\; E^n = -F^{0n},\; B^n = - \epsilon^{njk} F^{jk}/2$. Since we are working in lowest order only, there is no need of introducing matching coefficients in the Lagrangian (3.1).
We go to the rest frame of the bound state where $K^\mu = (-|E|,{\bf 0})$. Note that the rest masses are dropped in NRQCD, we will also suppress the $K$ in our notation from now on. The perturbation theory is performed in radiation gauge $\nabla {\bf A}=0$, the gauge best suited for treating bound states. From now on we distinguish between Coulomb gluons $A^0$ denoted by dashed lines, and transverse gluons ${\bf A}$ denoted by curly lines. 

We first solve the bound state equation in Fig. 4.
The leading contribution to the kernel $\tilde{I}_{Q\bar{Q},Q\bar{Q}}$ which appears in the first term on the rhs is given by the exchange of one Coulomb gluon between $Q$ and $\bar{Q}$ which leads to a static kernel:

\begin{equation} \tilde{I}_{Q\bar{Q},Q\bar{Q}} = \frac{i(2\pi)^4 g^2(\lambda^a_{\alpha \beta}/2)(\lambda^a_{\gamma \delta}/2)}{|{\bf p}-{\bf p'}|^2}. \end{equation}
The corresponding contribution with a transverse gluon contains additional factors of ${\bf p}/m$ at the vertices. It is therefore suppressed if we consider only soft external momenta. Internal momenta are also essentially soft due to the non relativistic wave function given below. The last three contributions in Fig. 4 are also of higher order. In lowest order Fig. 4 therefore gives the familiar  non relativistic Bethe Salpeter equation in ladder approximation and radiation gauge. Because the kernel is static, the only dependence on the relative energy variable $p^0$ on the rhs is in the propagators. One can therefore integrate over  $p^0$ on both sides and derive the Schr\"odinger equation for the momentum space wave function $\tilde{\psi }_{Q\bar{Q}}({ \bf p}) \equiv (2\pi )^{-5/2} \int \tilde{\psi }_{Q\bar{Q}}({ \bf p},p^0)dp^0$ in the usual way. After this has been solved, the dependence on the relative energy  can be recovered by going back to the original equation. The well known solution then reads

\begin{equation} (\tilde{\psi }_{Q\bar{Q}})_{nlm}({ \bf p},p^0) = \frac{i(2\pi)^{3/2}(|E_n|+{\bf p}_1^2/2m_1 + {\bf p}_2^2/2m_2 ) \tilde{\psi }_{nlm}({\bf p})}{( p^0 -\eta _1 |E_n|- {\bf p}_1^2/2m_1  + i\epsilon) (p^0 +\eta _2 |E_n|+ {\bf p}_2^2/2m_2  - i\epsilon)}. \end{equation}
The spin wave functions factorize. $E_n=- (C_F\alpha _s)^2\mu/(2n^2)$, with $C_F=(N^2-1)/(2N) = 4/3$ the Casimir operator in the fundamental representation, is the energy eigenvalue, and $\tilde{\psi }_{nlm}({\bf p})= \tilde{\psi }_{nl}(|{\bf p}|)Y_{lm}(\hat{{\bf p}})$ the normalized Schr\"odinger wave function in momentum space. The normalization has been chosen such that $\int \tilde{\psi }({\bf p},p^0) dp^0 =(2\pi )^{5/2}\tilde{\psi }({\bf p})$, which means that $\psi _{Q\bar{Q}}$ in (2.1) becomes the correctly normalized Schr\"odinger wave function if $x_1^0 = x_2^0$. The Bohr radius is $a=1/\sqrt{2\mu |E_1|}$. For the lowest states one has (it is convenient to write them in terms of the energy $E_n$ instead of the Bohr radius) 

\begin{eqnarray} 
\tilde{\psi }_{10}(|{\bf p}|) & = & \frac{4\sqrt{2}(2\mu |E_1|)^{5/4}}{\sqrt{\pi }({\bf p}^2+2\mu |E_1|)^2},\nonumber\\  
\tilde{\psi }_{20}(|{\bf p}|) & = & \frac{8\sqrt{2} (2\mu |E_2|)^{5/4}(|{\bf p}|^2 - 2\mu |E_2|)}{\sqrt{\pi }(|{\bf p}|^2+2\mu |E_2|)^3},\\
\tilde{\psi }_{21}(|{\bf p}|) & = &\frac{i16\sqrt{2}(2\mu |E_2|)^{7/4}|{\bf p}|}{\sqrt{3 \pi }(|{\bf p}|^2+2\mu |E_2|)^3}.\nonumber
\end{eqnarray}
Having solved the bound state equation in Fig. 4 in lowest order, one can insert the solution, together with the leading approximations for the propagators, the vertex functions, and the kernels, into the rhs of Fig. 5, dropping the higher order contributions of the second line. The power counting in the coupling constant $g$ requires some care, because there are not only the explicit factors $g$ in the vertices, but also factors $p/m \sim g^2$ in the coupling of transverse gluons to quarks, as well as momentum factors in the three gluon vertex. Taking this into account, the leading contribution to $\tilde{\psi }_{QA^0\bar{Q}}$ is then given by Fig. 6, the leading contribution to $\tilde{\psi }_{Q{\bf A}\bar{Q}}$ by Fig. 7.

\begin{figure}[htb]
\begin{center}
\epsfysize 2.7cm
\epsfbox{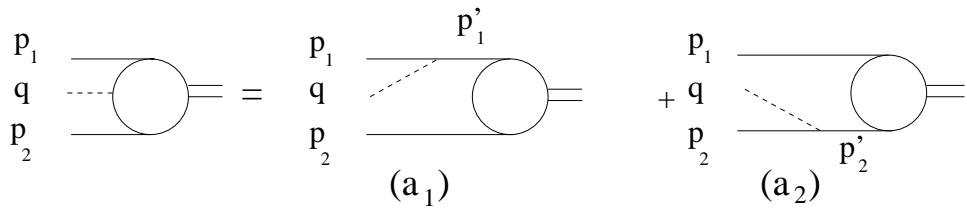}
\end{center} 
\caption{ \it The leading contributions to $\tilde{\psi }_{QA^0\bar{Q}}$.}
\end{figure}    

\begin{figure}[htb]
\begin{center}
\epsfysize 2.7cm
\epsfbox{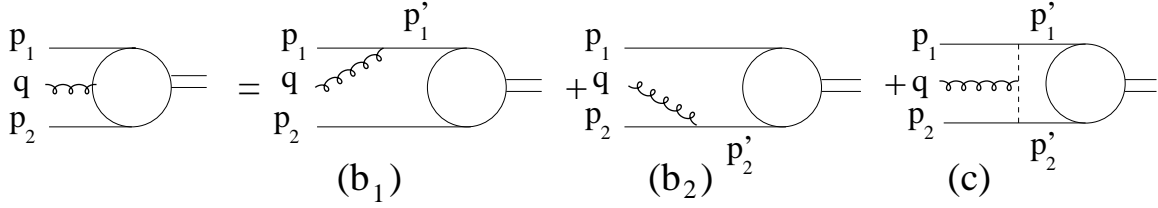}
\end{center}
\caption{ \it The leading contributions to $\tilde{\psi }_{Q{\bf A}\bar{Q}}$.}
\end{figure}    

In Fig. 7 c) the kernel $\tilde{I}_{Q{\bf A}\bar{Q},Q\bar{Q}}$ has been replaced by its lowest order contribution, where the transverse gluon couples to two Coulomb gluons, while the latter couple to the quark and antiquark, respectively. For the reasons mentioned above, this graph, although of higher order in the coupling constant, contributes to the same order as the tree graphs in Fig. 7 b), because the direct coupling of the transverse gluon to the quark or antiquark in 7 b) is suppressed by a factor $p/m \approx g^2$. For the  Coulomb gluon the corresponding graph does not exist, because there is no coupling of three Coulomb gluons. There is no non abelian contribution in leading order in Fig. 6.

We can now insert the $Q\bar{Q}$ wave function (3.3) into the rhs of Fig. 6 and Fig. 7, and calculate the wave functions on the lhs. 
We start with the tree graphs Fig. 6 a) and Fig. 7 b). They can be simply written down because there is no integration over internal momenta. We are interested in the equal time wave function, therefore we first integrate over $p^0$ (or, more simply, over $p_2^0$ in $a_1),b_1)$ and over $p^0_1$ in $a_2),b_2))$ with the residue theorem. There are three denominators, one arising from the external $Q$ or $\bar{Q}$ propagator and two from the $Q\bar{Q}$ wave function. For $a_1),b_1)$ e.g. one gets 

\begin{eqnarray} & & \int \frac{dp^0_2}{(p^0_2+q^0 + |E_n| + {\bf p}_1^2/2m_1 - i\epsilon) (p^0_2 + |E_n| + {{\bf p}'}_1^2/2m_1 - i\epsilon) (p^0_2 - {\bf p}_2^2/2m_2 + i\epsilon)}\nonumber\\ 
& = & - \; \frac{2\pi i}{(q^0 + |E_n| + {\bf p}_1^2/2m_1 + {\bf p}_2^2/2m_2- i\epsilon) (|E_n| + {{\bf p}'}_1^2/2m_1 + {\bf p}_2^2/2m_2)}. \end{eqnarray}
Next we have to perform the integration over $q^0$. 
For the Coulomb gluon the propagator is independent of $q^0$, one thus cannot simply perform the $q^0$-integration, because the denominator in (3.5) contains only one power of $q^0$. We therefore choose an infinitesimal relative time $t$ of the gluon with respect to the quark and antiquark. The Fourier transform then becomes 

\begin{equation} \int \frac{e^{-iq^0 t}dq^0}{(q^0 + |E_n| + {\bf p}_1^2/2m_1 + {\bf p}_2^2/2m_2- i\epsilon)} = 2\pi i \Theta (-t) \mbox{\quad for $t$ infinitesimal}. \end{equation}
We will comment this result below. Finally we add the contribution of Fig. 6 $a_2)$  and divide by $(2\pi)^5$ which gives the correctly normalized equal time wave function $\tilde{\psi }({\bf p},{\bf q})$ in momentum space. This results in 

\begin{equation} \tilde{\psi }_{QA^0\bar{Q}}({\bf p},{\bf q})
= \frac{g \sqrt{C_F} \Theta (-t)}{(2\pi)^{3/2}|{\bf q}|^2}\left( \tilde{\psi }(-{\bf p}_2) -
\tilde{\psi }({\bf p}_1)\right) . \end{equation}
For a transverse gluon  the propagator provides  enough powers of $q^0$ in the denominator to make the equal time wave function well defined. The result from the graphs Fig. 7 $b_1),b_2)$ becomes

\begin{eqnarray} (\tilde{\psi }_{Q{\bf A}\bar{Q}})^{(b)}({\bf p},{\bf q})
& = & \frac{g \sqrt{C_F} }{(2\pi )^{3/2} 2 |{\bf q}| \Big( |{\bf q}| + |E_n|  + {\bf p}_1^2/2 m_1+ {\bf p}_2^2/2 m_2\Big) } \times \nonumber\\ 
& & \left(\frac{[({\bf p}_1)_T +i{\bf q} \times {\bf s}_1 ]\tilde{\psi }(-{\bf p}_2)}{m_1}- 
\frac{[({\bf p}_2)_T +i{\bf q} \times {\bf s}_2 ]\tilde{\psi }({\bf p}_1)}{m_2}\right).\end{eqnarray}
Here ${\bf p}_T \equiv {\bf p} - ({\bf pq}){\bf q}/{\bf q}^2$ denotes the transverse component of ${\bf p}$.

We finally come to the non abelian graph Fig. 7 c). This  involves an integration over the internal four momentum $p'$. Performing first the integrations over $p'^0,p^0$, and $q^0$ gives the equal time wave function

\begin{eqnarray} (\tilde{\psi }_{Q{\bf A}\bar{Q}})^{(c)}({\bf p},{\bf q}) & = & \frac{g^3N\sqrt{C_F}}{(2\pi)^{9/2}2 |{\bf q}| \Big( |{\bf q}| + |E_n| + {\bf p}_1^2/2m_1 + {\bf p}_2^2/2m_2 \Big) } \times \nonumber\\
& & \int \frac{({\bf p}-{\bf p}')_T\tilde{\psi }({\bf p}')d^3 p'}{|{\bf p}-{\bf p'} - \eta _1 {\bf q}|^2 |{\bf p}-{\bf p}' + \eta _2 {\bf q}|^2}. \end{eqnarray}
We come to the physical interpretation of the above results. First we have to recall that there are three different scales in NRQCD: Hard ($\sim m )\gg$ soft $(\sim \alpha _s m )\gg$ ultrasoft $(\sim \alpha _s^2 m)$. Quark masses are hard, quark momenta soft, and quark kinetic energies as well as binding energies are ultrasoft. Since we used NRQCD, the momenta $p_1,p_2,q$ in the $QA\bar{Q}$ wave functions (3.7) - (3.9) must also be soft. In fact these expressions are dominated by soft momenta.

The graphs in Fig. 6 describe the Coulomb field moving along with the quark and antiquark. It is easy to perform the Fourier transform of (3.7) with respect to ${\bf p}$ and ${\bf q}$. It becomes 

\begin{equation} \psi _{QA^0\bar{Q}}
= \frac{g \sqrt{C_F}\Theta (-t)}{4\pi}\psi ({\bf r})\left( \frac{1}{|{\bf x}_3 - {\bf x}_1|} -  \frac{1}{|{\bf x}_3 - {\bf x}_2|}\right). \end{equation}
One recognizes the product of the $Q\bar{Q}$-wave function with the Coulomb potential generated by the $Q$ and the  $\bar{Q}$ source. (In QED one has, of course, to replace $g\sqrt{C_F}$ by $e$.) The appearance of the step function $\Theta (-t)$ is due to the time ordering prescription in the wave function (2.2). For $t>0$ the gluon field $A^0(x_3)$ has to stand to the left. One can express it in terms of the quark charge densities, and the latter annihilate the vacuum in lowest order.

The graphs in Fig. 7 b)  describe the fields which arise because the quarks are moving. For large $|{\bf q}|$ the spin independent contributions of the quark and antiquark are those of the Coulomb contributions (3.7) for $t<0$, multiplied with $(1/2)({\bf p}_j)_T/m_j$. With the exception of the factor $1/2$ in front, this is exactly what one would expect from classical electrodynamics. The appearance of this factor can be understood as follows. In classical radiation problems one uses the retarded Green function. In field theory, like in our definition of the wave function  (2.2), on uses, however, the time ordered product. If we had used the retarded product instead, the denominator of the transverse gluon propagator would have been changed from $q^2 + i\epsilon $ to $(q^0 + i\epsilon )^2 - |{\bf q}|^2$, such that both poles in the complex $q^0$-plane would lie in the lower half plane. In this case one would simply close the contour around the remaining $q^0$-pole from the quark propagator in the upper half plane and would end up with the replacement

\begin{equation} \frac{1}{2 |{\bf q}|\Big( |{\bf q}| + |E_n|+{\bf p}_1^2/2m_1 + {\bf p}_2^2/2m_2 \Big) } \Rightarrow  
\frac{1}{|{\bf q}|^2 -\Big( |E_n|+{\bf p}_1^2/2m_1 + {\bf p}_2^2/2m_2-i\epsilon \Big) ^2},\end{equation}
removing the factor $1/2$ for large $|{\bf q}|$ in (3.8). The use of the advanced propagator would, on the other hand, give a vanishing result. This shows that one has to define the equal time wave function very carefully. 

In the following we will stick to our definition with the time ordered product. If one defines the equal time wave function for $A^0$ as the average of the expressions for positive and negative infinitesimal $t$, the replacement $\Theta(-t) \Rightarrow 1/2$ leads to the correct relation between $A^0$ and ${\bf A}$ for large $|{\bf q}|$. For smaller $|{\bf q}|$ the result for ${\bf A}$ is modified, the denominator is less singular for $|{\bf q}|\rightarrow 0$. This is, of course, due to the fact that the quarks are not free, there is an additional  long range radiation field. Fig. 7 c) finally arises from the glue moving along with the quarks, it has no simple classical analog.

The expressions for the transverse gluon can be simplified. From the denominators in (3.8) and (3.9) it is obvious that these expressions are dominated by ultrasoft $q$. If we thus restrict to ultrasoft $q$ and remember that $p_1$ and $p_2$ are soft, we may approximate (3.8) by using ${\bf p}_1={\bf p} - \eta _1 {\bf q} \approx {\bf p} , \; -{\bf p}_2 = {\bf p} + \eta _2 {\bf q} \approx {\bf p}$, and ${\bf p}_1^2/2m_1 + {\bf p}_2^2/2m_2 = {\bf p}^2/2\mu + {\bf q}^2/2M \approx {\bf p}^2/2\mu $, and get

\begin{equation} (\tilde{\psi }_{Q{\bf A}\bar{Q}})^{(b)}({\bf p},{\bf q})
= \frac{g \sqrt{C_F} {\bf p}_T \tilde{\psi }({\bf p})}{(2\pi )^{3/2} 2 \mu |{\bf q}| \Big( |{\bf q}| + |E_n|  + {\bf p^2}/2 \mu \Big)  } .
\end{equation}
We dropped the spin dependent term because it is suppressed for ultrasoft $q$.

We next discuss the non abelian contribution (3.9).  For  ultrasoft $q$ one may again simplify the kinetic energies  in the denominator, furthermore one can drop the ${\bf q}$ in the integrand.  The integral can then be easily performed for a given $\tilde{\psi}({\bf p}')$, e.g. by introducing a Feynman parameter. We give the results for the three lowest states. 

\begin{eqnarray} (\tilde{\psi }_{Q{\bf A}\bar{Q}})_{100}^{(c)}({\bf p},{\bf q}) & = & \frac{g^3 N\sqrt{C_F}  (2\mu |E_1|)^{3/4}{\bf p}_T }{ (2\pi )^3 |{\bf q}| \Big( |{\bf q}| + |E_1|  + {\bf p}^2/2 \mu  \Big) ({\bf p}^2 + 2\mu |E_1|)^2}Y_{00}, \\
(\tilde{\psi }_{Q{\bf A}\bar{Q}})_{200}^{(c)}({\bf p},{\bf q}) 
& = & \frac{g^3 N\sqrt{C_F}  (2\mu |E_2|)^{3/4} ({\bf p}^2-3\cdot2\mu |E_2|){\bf p}_T}{(2\pi )^3|{\bf q}|\Big( |{\bf q}| + |E_2| +  {\bf p}^2/2 \mu  \Big)({\bf p}^2 + 2\mu |E_2|)^3}Y_{00}.\end{eqnarray} 
The result for the $P$-state consists of two parts:

\begin{eqnarray} 
(\tilde{\psi }_{Q{\bf A}\bar{Q}})_{21m}^{(c)}({\bf p},{\bf q}) 
&  = &  
 \frac{i4g^3 N\sqrt{C_F}   (2\mu |E_2|)^{5/4}}{(2\pi )^3 \sqrt{3}|{\bf q}|\Big( |{\bf q}| + |E_2|  + {\bf p}^2/2 \mu  \Big) } \times \nonumber \\
& &   \bigg( \frac{{\bf p}_T|{\bf p}|Y_{1m}(\hat{\bf p})}{({\bf p}^2 + 2\mu |E_2|)^3} 
- \frac{1}{4} \; \frac{({\bf e}_k)_T Y_{1m}({\bf e}_k)}{({\bf p}^2 + 2\mu |E_2|)^2} \bigg),
\end{eqnarray}
where ${\bf e}_k$ denote the three Cartesian unit vectors and $({\bf e}_k)_T$ their transverse parts. 

For the ground state, as well as for the first term of the $P$-state, we may express the result directly in terms of the wave function and compare it with the result of Fig. 7 b)  in (3.12). The non abelian contribution (3.13) for the ground state is $N/(2C_F)=9/8$ times the result in (3.12), while for the $P$-state the first contribution in (3.15) is $N/C_F=9/4$ times the corresponding result in (3.12). Thus numerically the non abelian loop graph is more important than the tree graphs. 

In momentum space the leading order of the transverse electrical field ${\bf E}_T=-\partial ^0{\bf A}$ and the magnetic field ${\bf B} = \nabla \times {\bf A}$ are obtained by replacing $\nabla \Rightarrow i{\bf q}$ and $\partial ^0 \rightarrow -iq_0 \Rightarrow -i|{\bf q}|$ and applying these factors in the expressions for ${\bf A}$. The last replacement is due to the fact that the $q^0$-integration was performed by closing the contour around the pole at $q^0 = |{\bf q}| -i\epsilon $. The longitudinal electric field ${\bf E}_L = -\nabla A^0$ can be  obtained from (3.7) by multiplying with $-i{\bf q}$, or directly in position space from (3.10).

\setcounter{equation}{0}\addtocounter{saveeqn}{1}%

\section{The $Q{\bf A}\bar{Q}$ wave function of the $P$-state at $r=0$}

A quantity of particular interest is the amplitude for the special case of quarks with identical flavor at zero distance in the $Q{\bf A}\bar{Q}$ component of the $P$-state. This component, although suppressed, contributes to annihilation decays to the same order as the $Q\bar{Q}$ annihilation in the leading $Q\bar{Q}$ component. The reason is that this annihilation is also suppressed, because the wave function at the origin vanishes for $P$-states. On the other hand the $Q\bar{Q}$ in the $QA\bar{Q}$ wave function can be in an $S$-state. In one loop order one finds that logarithmic divergences are canceled by adding up both contributions. This has been discussed in detail by Bodwin, Braaten and Lepage \cite{BBL}. Let us therefore consider the wave function for quarks at zero distance

\begin{equation}\tilde{\psi }_{Q{\bf A} \bar{Q}}({\bf q})_m \equiv (2\pi )^{-3/2}\int \tilde{\psi }_{Q {\bf A}\bar{Q}}({\bf p},{\bf q})_{21m}d^3p. \end{equation} 
The calculation is simple if we restrict to the dominating region where $q$ is ultrasoft. From (3.12), (3.15) we find

\begin{eqnarray} \tilde{\psi }_{Q{\bf A} \bar{Q}}({\bf q})_m & = &
\frac{ig\sqrt{C_F}(m|E_2|)^{5/4}({\bf e}_k)_TY_{1m}({\bf e}_k)}{3\sqrt{6}\pi ^{3/2} |{\bf q}| \left( \sqrt{m(|{\bf q}| + |E_2|)} + \sqrt{m|E_2|} \right) ^3}\times \nonumber\\
& & \left( 3 \sqrt{m(|{\bf q}| + |E_2|)} + \sqrt{m|E_2|} - \frac{2N}{C_F} \sqrt{m|E_2|} \right) .  \end{eqnarray}
The first two terms in the second line originate from the graphs in Fig. 7 b) and the resulting contribution (3.12), the third one is a sum of $\Big( N/C_F\Big) \; \Big\{ 3 \sqrt{m(|{\bf q}| + |E_2|) } +  \sqrt{m|E_2|}\Big\} $ from the first part of (3.15) and of $-\Big( 3N/C_F \; \Big) \Big\{  \sqrt{m(|{\bf q}| + |E_2|)} + \sqrt{m|E_2|}\Big\} $ from the second part. 

Our result (4.2) shows a remarkable feature. Recalling that $2N/C_F = 9/2$, one finds that the non abelian term dominates the abelian one for small $|{\bf q}|$ and makes the second line of (4.2) negative. For $|{\bf q}|=(13/36)|E_2|$ the bracket vanishes, for larger $|{\bf q}|$ the positive abelian term dominates. 

The wave function (4.2) determines the amplitude for annihilation of the two quarks in the color octet state into gluons which subsequently hadronize. As one might have expected, (4.2) depends on the gluon momentum ${\bf q}$, as well as on the relative orientation between the gluon and the angular momentum of the $P$-state. It is, however, independent of the quark spins. The $Q\bar{Q}$ wave function has entered in a complicated way. It is not possible to express the result simply by the derivative $R'(0)$ of the radial wave function at the origin.

\setcounter{equation}{0}\addtocounter{saveeqn}{1}%

\section{Conclusions}

In the foregoing we derived some results which may appear rather unexpected at first sight. One of them is the discontinuity of the wave function  in the relative time of the Coulomb gluon with respect to the quarks. The other one is the relation between transverse and Coulomb gluon. Both results could, however, be understood, the point is the use of different Green functions (retarded versus time ordered) in classical radiation theory and in field theory. 

One may wonder whether some of these features could be artifacts of the gauge or of the perturbative treatment. We are confident that this is not the case. One could, of course, choose manifestly gauge invariant wave functions by connecting $Q\bar{Q}$ or $QA^\mu \bar{Q}$, respectively, with path ordered strings. The $Q\bar{Q}$ wave function would now contain already some glue in it due to the presence of the string. This glue along the straight string is, however, not an additional degree of freedom. Everything is fixed by the positions of the quark and the antiquark. The essential feature of a $QA^\mu \bar{Q}$ wave function is not the presence of some glue, but the presence of an additional degree of freedom carried by the gluon.

The strings would not contribute in lowest order perturbation theory. A consideration of higher order perturbative contributions within the approach presented here, whether with or without strings, is certainly possible. It will not alter the essentials of our results. Our method could also be extended to obtain wave functions for multi gluon admixtures which arise in higher orders. 

A systematic treatment of non perturbative effects is beyond the present investigation. We emphasize, however, that  our whole approach can be immediately generalized to the case where we replace the kernel (3.2) by some other static kernel, say the Fourier transform of a phenomenological static potential involving confinement. This will, in turn, modify the $Q\bar{Q}$ Schr\"odinger wave function and the subsequent results. For phenomenological applications such an approach appears quite legitimate, it has been familiar and successful in other fields of quarkonium physics. 

The application of our results to production or decays of heavy quarkonia goes beyond the present investigation. In the framework of NRQCD, hard annihilation processes are described in lowest order by four fermion terms of dimension six. They were not included in our Lagrangian because they are of order $1/m^2$ and thus not important for the calculation of the wave function. While the hard subprocess of $S$-wave annihilation is simple and well understood, the complicated dependence of our wave function on the gluon momentum and the magnetic quantum number of the $P$-state require some additional work on this topic. While we know that logarithmic divergences which show up in higher orders cancel exactly against corresponding divergences in the $P$-wave annihilation of the $Q\bar{Q}$ component, there is no information about the finite part up to now. This information can be obtained from our explicit wave function. Due to the compensation between the two contributions in (4.2) and the change in sign between smaller and larger $|{\bf q}|$, suppressions compared to naive expectations may be expected. The results will be presented elsewhere. \\

\noindent
{\bf Acknowledgement}: I thank D. Melikhov and N. Brambilla for reading the manuscript and for valuable suggestions.

\setcounter{equation}{0}\addtocounter{saveeqn}{1}%

\begin{appendix}
\section{Decomposition of Green functions}

In this appendix we prove the correctness of the decompositions in Fig. 2 (the proof for Fig. 3 is analogous) by showing that every graph on the lhs appears exactly once on the rhs.
Consider an arbitrary graph contributing to the lhs. 

1) If the graph is disconnected it uniquely belongs to the first term on the rhs of the first line of Fig. 2. 

Consider a connected graph next. 

2) If the graph cannot be separated by a $Q\bar{Q}$ or a $QA\bar{Q}$ internal cut, it belongs to $I_{Q\bar{Q},Q\bar{Q}}$ plus external lines. It then belongs to the second term on the rhs,  where the disconnected part of the Green function appears. 

Let's next take the case where the graph can be cut by an internal $Q\bar{Q}$ or $QA\bar{Q}$ cut. The propagators are always taken as part of the right hand graph. We first construct a unique preliminary cut $C_p$  as follows: Follow the quark line from the left until one comes to an internal propagator where the graph can be cut in the above way. If there are several possibilities to do so, let $C_p$ be  the cut where the antiquark line is cut as far to the left as possible. 

3) If $C_p$ is a $Q\bar{Q}$ cut we take it as the final cut $C$. In this case the graph belongs to the second term on the rhs of Fig. 2 with a connected contribution to the Green function.

If $C_p$ is a $QA\bar{Q}$ cut there are two alternatives: 

4) It may be possible that one can transform the cut by keeping it's position at the $Q$ (or $\bar{Q}$) line, while moving the position of the cut at the $\bar{Q}$ (or $Q$) to the right, such that one  obtains a $Q\bar{Q}$ cut. This happens if one can shift the cut to the right by moving it over a vertex part, such that it does no longer cut the gluon line. If this procedure is possible at all, it is unique. In this case we take this $Q\bar{Q}$ cut as the final $C$. Note that the graph left to the cut is a contribution to $\tilde{I}_{Q\bar{Q},Q\bar{Q}}$; it is not allowed to cut the graph by the original $C_p$, because it would cut an {\em external} $Q$ or $\bar{Q}$ line now. Again we get a contribution to the second term on the rhs of fig. 2.

5) If the above manipulation is not possible, we take the original $QA\bar{Q}$ cut $C_p$ as the final $C$. In this case the graph on the lhs is a part of $\tilde{I}_{Q\bar{Q},QA\bar{Q}}$, while the one on the rhs is a part of the totally connected Green function $\tilde{G}_{QA\bar{Q},Q\bar{Q}}^{conn}$. Contributions to the first two terms on the rhs of Fig. 1 cannot appear by construction because our prescription would have moved the corresponding cut to the right. 

We have thus shown that every graph on the lhs also appears on the rhs. In order to proof that it belongs exactly to one term on the rhs we will next
 show that it is not possible to move the cut $C$ to the right, producing thereby another admissible cut $C_r$, which would lead to a different decomposition.  (By construction it is certainly not possible to move the cut to the left).

Consider first the case that $C$ is a $Q\bar{Q}$ cut. If we move it to the right, crossing a vertex both at the $Q$ line as well as at the $\bar{Q}$ line (irrespective whether we arrive at an admissible cut), the remaining graph on the left could be cut by the original $C$ and thus would not belong to an irreducible kernel. If we move the cut only over one vertex, such that it becomes a $QA\bar{Q}$ cut, the graph on the lhs would be inacceptable because of rule b) in the definition of the kernels in sect. 2. 

Consider next the case that $C$ is a $QA\bar{Q}$ cut. If we move it to the right by crossing a vertex both at the $Q$ line as well as at the $\bar{Q}$ line (irrespective whether we arrive at an admissable cut), the remaining graph on the left could be cut by the original $C$ and thus would not belong to an irreducible kernel. Rule d) of sect. 2 applies in the case that the gluon line which was cut by $C$ is also cut by $C_r$ and thus would become an external line. 

If we move the cut only along the $Q$ (or the $\bar{Q}$) line while keeping the position at the $\bar{Q}$ (or the $Q$) line we may either obtain a $Q\bar{Q}$ cut or a $QA\bar{Q}$ cut (or an inadmissible cut). In the first case the kernel to the left is forbidden by rule b), in the second case by c).

The somewhat complicated definition of the kernels $\tilde{I}_{Q\bar{Q},QA\bar{Q}}$, $\tilde{I}_{QA\bar{Q},QA\bar{Q}}$ with respect to external lines at the rhs was necessary to guarantee the uniqueness of the decomposition. In Fig. 8 we show some graphs which do {\em not} belong to $\tilde{I}_{Q\bar{Q},Q\bar{Q}}$ or to $\tilde{I}_{Q\bar{Q},QA\bar{Q}}$, together with the cuts  and the rule of sect. 2 which prevents this.

\begin{figure}[htb]
\begin{center}
\epsfysize 3.0cm
\epsfbox{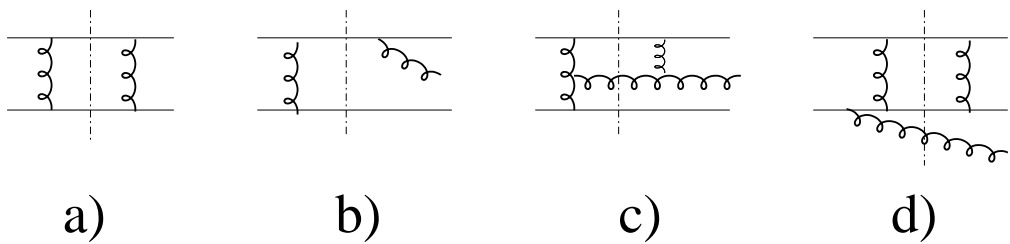}
\end{center}
\caption{ \it Some reducible kernels.}
\end{figure}    
 
\end{appendix}

\begin{thebibliography}{99}


\bibitem{BBL} G. T. Bodwin, E. Braaten, G. P. Lepage, Phys. Rev. D {\bf 46}, R1914
(1992);\\ 
with T. C. Yuan, Phys. Rev. D {\bf 46}, R3703 (1992);\\
 Phys. Rev. D {\bf 51}, 1125 (1995), Phys. Rev. D {\bf 55}, 5853(E) (1997).

\bibitem{NRQCD} W. E. Caswell, G. P. Lepage, Phys. Lett. B {\bf 167}, 437 (1986).

\end{thebibliography}
\end{document}